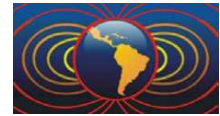

# ANALISIS ESPECTRAL DE PERIODO CORTO DEL INDICE K REGISTRADO EN EL OBSERVATORIO GEOMAGNETICO DE HUANCAYO

Domingo Rosales[1*], Erick Vidal[1]

[1] Observatorio Geomagnético de Huancayo – Instituto Geofísico del Perú, Huancayo, Perú.
[*] e-mail: domingo.igp@gmail.com.

**ABSTRACT**

Wavelet spectral analysis is applied to the daily indices K (SK) registered in Huancayo geomagnetic observatory from 2000.0 to 2015.0, it has become possible to identify predominant periodic components with a confidence level of 95%. In particular we have investigated the periodicity of 27.0 days and its 13.5, 9.0 and 6.8 harmonics, related to the fundamental period of solar rotation effects. A special analysis registered on periods of 59.0 days might be related to the double period of 29.53 days corresponding to the synodic month. The manifestation of the periods of 182.6, 365 and 730 days corresponding to the harmonics of the seasonal variation of the annual cycle is also observed. An important observation is that the global predominant periods registered 6.8 and 9.0 days, in the years 2009, 2013 and 2014 its effect is too small or null, those periods corresponds to the minimum and maximum solar activity. On the other hand, it is observed that the order of predominance of the harmonic period solar rotation are different for each solar cycle (Solar Cycle 23: 9.0, 13.5, 27.0 and 6.8 days; solar cycle 24: 27.0, 13.5, 9.0 and 6.8 days), and it is weaker contribution in the solar cycle 24 with respect to the solar cycle 23.

**RESUMEN**

El análisis espectral por wavelet, es aplicado a los índices K diarios (SK) registrados en el observatorio geomagnético de Huancayo entre los años 2000.0 y 2015.0, ello ha hecho posible identificar componentes periódicas predominantes con un nivel de confianza del 95%. En particular se ha investigado la manifestación de la periodicidad de 27.0 días y sus armónicas de 13.5, 9.0 y 6.8 días, efectos relacionados con el periodo fundamental de rotación solar. Además, se hace un especial análisis sobre la manifestación del periodo registrado de 59.0 días que estaría relacionado con el periodo de 29.53 días que corresponde al mes sinódico. También se observa la manifestación de los periodos de 182.6, 365 y 730 días que corresponde a las armónicas de la variación estacional del ciclo anual. Una importante observación resalta el echo que los periodos predominantes registrados globalmente de 6.8 y 9.0 días, en los años 2009, 2013 y 2014 su presencia es poca o nula, periodos que corresponde al mínimo y máximo de actividad solar. Por otro lado, se observa que el orden de predominancia de las armónicas del periodo de rotación solar son distintas para cada ciclos solar (ciclo solar 23: 9.0, 13.5, 27.0 y 6.8 días; ciclo solar 24: 27.0, 13.5, 9.0 y 6.8 días), y que además es más débil su contribución en el ciclo solar 24 con respecto al ciclo solar 23.

**Palabras Clave**: Índice K, actividad geomagnética, geomagnetismo, análisis espectral wavelet.

**Introducción**

El índices K fue introducido por Bartels, como una medición de la variación de la actividad geomagnética registrado en un punto de la superficie terrestre, son números enteros con rango de 0 a 9; de 0 a 3 indica condiciones de calma y mayor a 5 indica condiciones de tormenta. Los índices K son calculados para cada observatorio. Es un índice cuasi-logarítmico derivado de la máxima fluctuación de la componente H, registrado con un magnetómetro durante un intervalo de 3 horas (Bartels *et al.*, 1939;

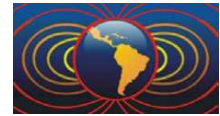

Menvielle, 2002). Históricamente, los índices K se obtenían mediante escalado manual a partir de los magnetogramas fotográficos. Con el avance del desarrollo electrónico y proceso computacional, actualmente se obtienen a partir de algoritmos computacionales que extraen la información de los magnetómetros digitales (Hattingh *et al.,* 1989; Menvielle *et al.*, 1995; Stankov *et al.*, 2010).

Los ciclos de actividad solar son caracterizados por cambios recurrentes en la actividad geomagnética, los cuales varían en función de de la magnitud del campo magnético solar, las manchas solares, agujeros coronales y las eyecciones de masa coronal (CME) y rotación solar, por lo que los niveles de actividad geomagnética durante el periodo de la fase del mínimo solar son considerablemente diferentes al periodo de la fase de máximo solar, y así mismo el efecto de la rotación solar también produce variaciones en la actividad geomagnética, por lo que diversos estudios dan cuenta que el periodo dominante recurrente de la actividad geomagnética es el periodo de 27.0 días que corresponde con el ciclo de rotación solar (Bartels, 1934; Ogg, 1946; Nevanlinna *et al.,* 2011; Rosales *et al.*, 2012).

Las amplitudes de la actividad geomagnética en latitudes altas, medias y bajas son distintas, siendo un caso especial la actividad geomagnética en la zona del ecuador geomagnético debido a la fuerte influencia del "electrochorro ecuatorial" que amplifica la actividad geomagnética, poniendo en el mismo orden de actividad a lo registrado en zonas de latitudes medias y altas. El observatorio geomagnético de Huancayo (lat. 12.04º S, long. 75.33º O, alt. 3314 m.s.n.m), se encuentra localizado dentro de la zona del ecuador geomagnético (año 2015.0: lat. mag. 2.28º S, long. mag. 2.65º O, Dip -0.24), teniéndose en cuenta que en Julio del año 2013 se registro el paso del ecuador geomagnético por el observatorio de Huancayo, desplazándose a una velocidad de 8.6 km/año, tal como fue pronosticado en el trabajo de D. Rosales (Rosales *et al.,* 2011). Actualmente el ecuador geomagnético se encuentra a 17 kilómetros al norte del observatorio de Huancayo (manteniéndose aun dentro de la influencia de la zona del electrochorro ecuatorial).

En este trabajo se usan el análisis espectral por wavelet, para determinar las variaciones de los índices de actividad geomagnética en Huancayo. Se analiza la actividad geomagnética debido al efecto solar entre el año 2000.0 al 2015.0. Además se hace un estudio de dos periodos muy significativos de actividad solar (2009) y (2013-2014) que corresponden a la mínima y máxima actividad solar respectivamente.

**Datos**

Los índices K (K9=600 nT) de Huancayo fueron obtenidos de ReseachGate (www.researchgate.net) para los años 2000.0 al 2015.0 (Rosales y Vidal, 2015a; Rosales y Vidal, 2015b; Rosales y Vidal, 2015c; Rosales y Vidal, 2015d). Son 15 años de datos con registro completo de los cuales los índices K diarios (SK) son extraídos para su análisis espectral.

**Análisis Wavelet**

El análisis wavelet es una herramienta muy potente para descomponer una serie de tiempo en el espacio tiempo – periodo. Su particular propiedad de esta técnica es que permite determinar periodos transitorios, cuasi-periodos y periodos que cambian en el tiempo. En esta investigación es aplicado el wavelet de Morlet $\varphi(t)$ consistente en una onda plana modulada por una Gausiana dada por:

$$\varphi(t) = t^{-1/4} e^{i\omega_0 t} e^{-t/2}$$

con frecuencia adimensional ($\omega_0 = 6$) y $t$ como parámetro temporal (Morlet *et al.,* 1982; Torrence and Compo, 1998).

El análisis wavelet es posible representar señales $f(t)$ por medio de una serie definida por:

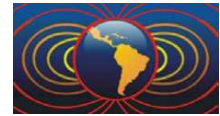

$$f(t) = \sum_{m=-\infty}^{\infty} \sum_{n=-\infty}^{\infty} c_n^m \varphi_n^m(t)$$

donde $\varphi_n^m(t) = \varphi(2^m t - n)$ es llamado el wavelet madre, y $c_n^m$ son los coeficientes.

**Resultados y discusión**

El espectro de potencia wavelet es aplicado a los índices K diarios (SK) (figura 1a), obteniéndose como resultado el espectro de potencia con niveles de contorno de $5^2, 20^2, 40^2, y\ 60^2\ SK^2$ (figura 1b). En el espectro de potencia global (figura 1c) se identifican los periodos de 6.8, 9.0, 13.5, 27.0, 59.0, 182.6, 365 y 730 días, con un nivel de confianza del 95%. Los periodos de 6.8, 9.0 y 13.5 días corresponden a las armónicas del periodo fundamental de 27.0 días, siendo los periodos mas dominantes de 9.0 y 6.8. Los periodos de 182.6, 365 y 730 días, corresponden a las armónicas del ciclo anual.

Las armónicas del periodo fundamental de rotación solar (27.0 días) están presente a lo largo de los 15 años de análisis, teniendo como periodos predominantes los de 9.0 y 6.8 días. Sin embargo durante el año 2009 el único periodo observable es el de 13.5 días. Durante el año 2013 los únicos periodos observables son de 5.4 (quinta armónica del periodo de 27 días) y 6.8 días. Y en el año 2014 el único periodo observable es el de 5.4 días, las demás armónicas del ciclo de rotación solar fundamental no se verifica. Por otro lado entre los años 2000.0 y 2009.0 (ciclo solar 23) es muy marcada la contribución de las armónicas del periodo de rotación solar, predominando los periodos en el siguiente orden: 9.0, 13.5, 27.0 y 6.8 días, en tanto que del año 2009.0 al 2015.0 (ciclo solar 24) los periodos predominantes tienen el siguiente orden: 27.0, 13.5, 9.0 y 6.8 días, además es muy débil su contribución con respecto al ciclo solar 23 (figura 1b).

Las armónicas correspondientes al ciclo anual a lo largo de los 15 años de análisis también presentan variaciones, es así que desde inicios del año 2009 hasta finales del año 2013 no es observable el ciclo anual ni sus armónicas, de igual forma gran parte del año 2000 no se registra el ciclo anual.

El efecto lunar en la actividad geomagnética diaria ha sido puesto en evidencia en muchos trabajos desde 1841 (Rosales *et al.*, 2013). También es sabido que la magnitud de este efecto es mayor en la región de América del Sur. Un detallado examen en los datos geomagnéticos de Huancayo para los años 1922 a 1939 fue realizado por Bartels y Johnston (1940), mostrando una marcada contribución lunar diurna y semidiurna. Sin embargo la Luna aparentemente parece no tener un efecto geomagnético apreciable de periodo mensual (29.53 días). Análisis armónicos mensuales y semi-mensuales de la componente H fueron realizados por diversos investigadores mostrando alguna evidencia, sin embargo la no plena evidencia se atribuye a la contaminación de los datos por las perturbaciones geomagnéticas, o que dicha aparente evidencia seria debido a otras causas distintas a la lunar como por ejemplo bandas subsidiarias del ciclo solar de 11 años (Stening, 1975; Střeštik, 1998). El mes sinódico es el intervalo entre dos sucesivas fase lunares (de Luna nueva a la siguiente Luna nueva) y refleja la posición de la Luna con respecto a la línea Sol-Tierra. Este mes sinódico es de 29.5305 días. En el análisis espectral global wavelet (figura 1 c) el periodo de 59.0 días es evidenciado con un nivel de confianza del 95%, que correspondería con el doble ciclo del mes sinódico ($59/2 = 29.5$ días). Aunque no se observa el ciclo de 29.5 y 14.8 días, el periodo registrado de 59.0 días seria una plena evidencia de la influencia de la sub-armónica del periodo principal de 29.53 días. Por otro lado tomando en cuenta que análisis espectrales aplicados sobre los índices Ap en un rango de 60 años evidencian periodos de 29.6 y 14.75 días (Střeštik, 1998), lo que implicaría que para registrar la evidencia de la influencia del mes sinódico y sus armónicas se requeriría mayor rango de datos a ser analizados.

**Conclusiones**

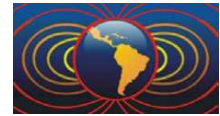

Se verifica la manifestación del periodo fundamental de rotación solar de 27.0 días y sus armónicas en los índices K diarios (SK). Así mismo en el inicio del ciclo solar 24 (año 2009) y el máximo solar (años 2013 y 2014) la manifestación del periodo de rotación solar es mínima o nula.

El orden de predominancia de las armónicas del periodo de rotación solar son distintas para cada ciclo solar (ciclo solar 23: 9.0, 13.5, 27.0 y 6.8 días; ciclo solar 24: 27.0, 13.5, 9.0 y 6.8 días), y demás se verifica que en el ciclo solar 24 son 7 veces mas débiles en su contribución respecto al ciclo solar 23, por lo que se concluye que el efecto de la actividad solar debido a la rotación solar es menos intensa respecto al ciclo solar 23.

Se verifica el periodo de 59.0 días a un nivel de confianza del 95%, periodo que correspondería con el doble ciclo del mes sinódico de 29.53 días, lo que seria una plena evidencia de la influencia de la sub-armónica del periodo principal del mes sinódico.

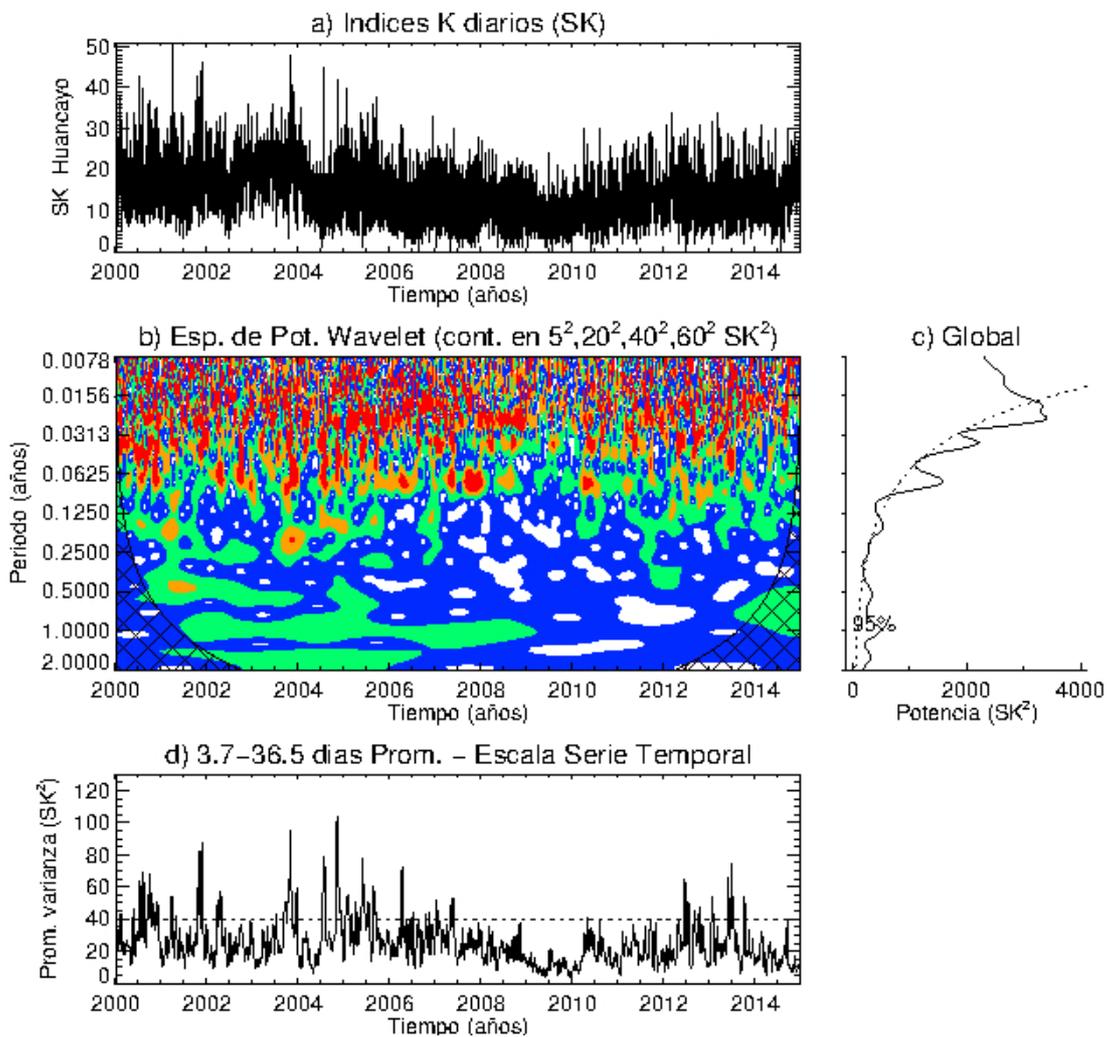

Figura 1 a) Índices K diarios (SK) desde el año 2000.0 al 2015.0. b) Espectro de potencia wavelet, con niveles de contorno de $\mathbf{5^2, 20^2, 40^2, y\ 60^2\ SK^2}$. c) Espectro de potencia wavelet global con un nivel de confianza del 95%. d) Varianza promedio entre 3.7 y 36.5 días.

**Referencias**

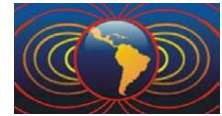